\documentclass[aps,pre,twocolumn,groupedaddress,showkeys]{revtex4-2}

\usepackage{amsmath,amssymb,amsfonts}
\usepackage{dcolumn}
\usepackage{graphicx}
\usepackage{orcidlink}
\usepackage{hyperref}

\begin{document}


\title{Controller for rejection of the first harmonic in a periodic signal with uncertain delay}


\author{Viktor Novi\v{c}enko}
\email[]{viktor.novicenko@tfai.vu.lt}
\homepage[]{http://www.itpa.lt/~novicenko/}
\affiliation{Institute of Theoretical Physics and Astronomy, Vilnius University, Vilnius, LT-10257 Lithuania}

\author{\v{S}ar\={u}nas Vaitekonis}
\email[]{vaitekonis@ftmc.lt}
\affiliation{Department of Nanoengineering, Center for Physical Sciences and Technology, Vilnius, LT-02300 Lithuania}


\date{\today}

\begin{abstract}
The plant (the system to be controlled) is disturbed by a periodic external force with a broad spectrum of Fourier harmonics. The first Fourier harmonic (sine-type signal) is assumed to be undesirable and should be removed by a control force, whereas the other harmonics should be preserved without distortion. Because the measured plant data have an unknown amount of time delay (dead time) and the sensitivity of the plant to external force is unknown, therefore the amplitude and phase of a sine-type control force are also unknown. Based on the internal model principle, we developed a feedback controller described as a linear time-invariant system that can remove the first harmonic from the plant's output by constantly adjusting its control force parameters. Using this controller, we aimed to further extend the capabilities of a newly developed high-speed, large-area rotational scanning atomic force microscopy. In such a scanning technique, the sample is rotated, and the tilt angle between the normal of the sample surface and the axis of rotation produces a parasitic first Fourier harmonic, which significantly limits the scanning area.
\end{abstract}

\keywords{Atomic force microscopy, uncertain delay, Internal model principle, Laplace transform, Linear time-invariant control, System of harmonic oscillators}

\maketitle


\section{\label{sec:intro}Introduction}

Rejecting a periodic disturbance acting on a controlled system has many engineering applications. Among these endless lists of applications, one can mention helicopter vibration reduction~\cite{Friedmann1995,Patt2005,Welsh2018}, reduction of side-by-side tower oscillations in wind turbine~\cite{Pamososuryo2024}, vibration compensation in hard disk drives~\cite{Yabui2019,Yabui2020}, wave cancellation in marine systems~\cite{Basturk2013}, active noise control~\cite{Elliott1992,Chandrasekar2006}, spacecraft vibration isolation~\cite{Chen2018,Shi2024}, suppression of torque ripple~\cite{Mosskull2024,Chen1990,Mandel2015} or cantilever beam~\cite{Yue2017} oscillations. As it is shown in Refs.~\cite{renzi204,Machado2024} such applications can have a significant economic impact.

The internal model principle (IMP)~\cite{conant1970,Francis1976,Wonham1976} is a fundamental mathematical component of many controllers that cancel periodic disturbances. The IMP states that the controller should contain a model that generate disturbance and reference signals. For example, in a well-known proportional-integral-derivative~(PID) controller designed to track a constant reference value, the main part is integral, because it produces a constant control force that generates a constant reference signal in the output of the close-loop system. While the role of proportional and derivative parts is to improve the stability and robustness of the close-loop system. In case of a periodic disturbance, the IMP-based controller should contain harmonic signal generators with the appropriate frequencies.

Control algorithms have many names that address various problems related to canceling periodic disturbances. One of the oldest methods is called a repetitive control (RC), which first appeared in Ref.~\cite{Inoue1981} and has been developed further by many researchers~\cite{Hara1988,Tomizuka1989,Escobar2005,Zhang2003}. It has a delay line equal to the period of the fundamental frequency. The objective of RC is to reject all harmonics in the known frequency periodic disturbance even if the number of harmonics is infinite (in the case of a discrete implementation, the number of harmonics is limited by the Nyquist frequency). RC can be generalized to a fractional-order RC~\cite{Fedele2017,Fedele2018} to improve its stability.

Another method closely related to RC is called resonant control (RSC), which has a sine-wave generator and can cancel single-tone disturbances of known frequency. As shown in Refs.~\cite{Escobar2005,Lu2010}, RC is equivalent to infinitely many RSCs connected in parallel with the frequencies integer multiple to the fundamental frequency. Such decomposition allows the generalization of the RC~\cite{WenzhouLu2013} designed for the faster rejection of specific harmonics~\cite{Lu2016} (for example, when the harmonic number is $4k\pm 1$ or $6k\pm 1$).

IMP-based adaptive feedforward cancellation (AFC)~\cite{Chen1990} is another family of methods for the suppression of periodic (or quasi-periodic in the case of multiple fundamental frequencies) disturbances having a finite number of harmonics. Although the AFC has non-linear equations, as shown in Refs.~\cite{Bodson1994,Messner1995,Bodson2004}, it is equivalent to RSC. Over the decades, many modifications of the AFC have been developed, such as working with unknown frequencies~\cite{Bodson1997} (also see the references in~\cite{Bodson2016}), removing a strictly positive real (SPR) condition~\cite{Wang2020}, and eliminating a negative impact due to a waterbed effect~\cite{Yabui2021}.

All the aforementioned methods continuously adapt amplitudes and phases of the harmonic components in the control force. This is especially evident for the AFC case, where differential equations are derived for the dynamics of the amplitudes and phases. In contrast, a discontinuous adaptation of the amplitudes and phases is used in a harmonic steady-state (HSS)/higher harmonic control (HHC) methods~\cite{Friedmann1995,Nygren1989,Knospe1996,Chandrasekar2006}. Among the modifications of HSS/HHC, one can mention time-domain HSS/HHC~\cite{Kamaldar2020,Kamaldar2020a}, which allows a faster convergence time.

A \textit{non-linear dynamics} community studies similar control problems (from a mathematical point of view) as mentioned above, yet usually without knowing and referring to the achievements of the \textit{control engineering} community. For example, RC has a transfer function (see, for instance, Eq.~(6) in Ref.~\cite{Escobar2005}) multiplicative inverse to a transfer function of extended delayed feedback control (EDFC)~\cite{soc94} used to stabilize an unstable periodic orbit of a known period but unknown profile. A similar situation happens with a method of harmonic oscillators (MHO)~\cite{Olyaei2015,AzimiOlyaei2018} and IMP-based RSCs connected in parallel. In fact, the derivation of the equivalence between the RC and infinitely many RSCs~\cite{Escobar2005,Lu2010} and the derivation of the equivalence between the EDFC and MHO~\cite{PyrHarmOsc2015} share the same mathematics. Yet, we should point out that both communities address different problems, that is, the cancellation for the control engineering community and the stabilization for the non-linear dynamics community. Nevertheless, each community can benefit from each other knowledge, and our paper is devoted to doing so.

In this work, we address the problem of cancellation of only the first harmonic of the periodic disturbance signal while all other harmonics, including the zeroth harmonic, should remain unchanged. This is the main difference between our approach and regular cancellation algorithms, such as RC, RSC, AFC, HSS/HHC. In addition, we assume that the plant has an unknown amount of time-delay. Therefore, our algorithm should automatically adjust the appropriate phase for a sine wave. The demand for our task appears naturally in experiments using rotational scanning Atomic Force Microscopy (AFM)~\cite{Ulcinas2017}, where the sample is rotated with a constant known angular frequency and the tip performs a circular trajectory. The surface of the sample is always tilted with respect to the rotation axis (see Figure~\ref{fig1}). Therefore, the parasitic first Fourier harmonic appears in the output signal. In a standard AFM, where a raster scanning trajectory is used, the typical solution to the tilt problem utilizes a PID controller. However, we can not expect a constant-height working mode for our circular trajectory due to the fast rotation speed and large circle radius. However, we can expect to avoid the tilt problem by filtering out the first harmonic in real-time. This allows us to scan larger areas, because there is no signal saturation once we go far away from the rotation axis. Therefore, we developed our controller. The successful experimental implementation of our controller was demonstrated recently in Ref.~\cite{VAITEKONIS2025117552}.

As a primary influence, we exploit MHO~\cite{Olyaei2015,AzimiOlyaei2018} and modify it to cancel the selective (in our case, the first) harmonics. A similar selective harmonic cancellation was analyzed in Ref.~\cite{Mattavelli2004}. However, here we deal with continuous signals (without the limiting Nyquist frequency) and a plant with an unknown time delay. Due to infinite harmonics, we obtain the delay differential-algebraic equations describing the controller dynamics. The main contributions of this article are as follows:
\begin{itemize}
  \item A linear time-invariant controller for rejecting the selected harmonic is designed as a set of specifically coupled harmonic oscillators (or RSCs) acting in parallel.
  \item Stability conditions are considered due to uncertain delay.
  \item The controller version containing infinite harmonic oscillators is obtained as a set of delay differential-algebraic equations.
\end{itemize}

The rest of this paper is organized as follows. In Section~\ref{sec:prob} we explain the experimental setup and formulate a mathematical problem in detail. While the solved mathematical problem is inspired by rotational scanning AFM, the presented controller can be implemented in any experiment with a similar mathematical model. In Section~\ref{subsec:cont} we describe the construction of our controller and analyze the stability questions. We also show a numerical simulation of the closed-loop system, which proves that the controller can successfully achieve this goal. Because the controller is a linear time-invariant system, Section~\ref{sec:block_scheme} presents the block diagram of the controller and the frequency response. Section~\ref{sec:infinite} is devoted to the most general version of the controller with infinite harmonics. Here, we discuss the advantages and disadvantages of an infinite harmonic number in the controller. Finally, we summarize the findings in Section~\ref{sec:conc}. Various time-demanding derivations have been moved to the appendices.

\section{\label{sec:prob}Problem formulation}

AFM is a scanning probe microscopy technique that gathers the sample surface data using a mechanical probe. Usually, this is achieved using a raster scanning technique when the tip scans the area of the sample line by line using a rectangular pattern. The roughness of the surface disturbs the tip of the microscope probe by moving it up and down from its resting state. Such disturbances are measured in real-time and, together with positional data of the scanner, serve as very high-resolution 3D topographic images of the surface. By indenting the probe into the sample and recording the applied force, which is proportional to the cantilever deflection, and the distance traveled by the probe one can also map various bio-mechanical properties and responses of soft matter down to the single-molecule level~\cite{Magazzu2023,Krieg2019}. In contrast to conventional raster scanning, which is used in most commercially available instruments, rotational scanning AFM~\cite{Ulcinas2017} uses a spindle to rotate the sample around the rotation axis at known angular frequency $\omega$. The tip is slowly mowed along the sample surface, usually starting from the center of rotation, therefore producing a spiral-type scanning path. Such a scanning technique resembles a phonograph where a stylus follows a groove on a rotating vinyl disk. This method has a significant advantage over raster scanning because the tip does not have to abruptly change the direction of its movement, therefore making it at least an order of magnitude faster compared to raster scanning and, as a consequence, allows scanning of much larger areas in the same amount of time.

\begin{figure}[h!]
\centering\includegraphics[width=0.85\columnwidth]{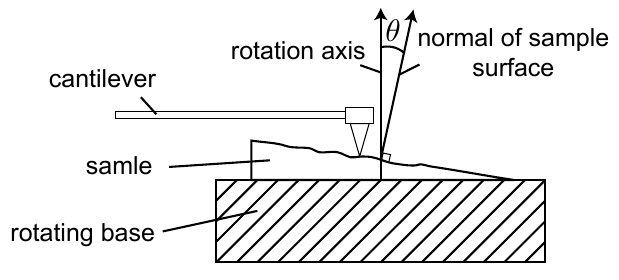}
\caption{\label{fig1} Schematic illustration of the tilt problem occurring in the rotational scanning.}
\end{figure}

Unfortunately, due to imperfections of mechanical machining, the normal of the sample surface is always tilted at some angle $\theta$ with respect to the spindle rotation axis (see Figure~\ref{fig1}). Therefore, the topography signal always contains the first Fourier harmonic, which does not provide relevant information about the sample surface features. Moreover, the amplitude of the first harmonic increases when the probe moves away from the center of rotation. This eventually completely saturates the topography signal and significantly limits the scanning area, as tip displacement detectors have a limited linear range of operation. One of the possible solutions to the tilt problem is to construct the feedback loop based on the real-time topography data, which moves the sample up and down using an additional actuator to eliminate the first Fourier harmonic in the output signal. Since the actuator is an electromechanical device with its own inertia and internal control loop, there is a delay between the command and the action. Moreover, this delay is unknown and may vary for different amplitudes of the sine-type signals. As a result, we end up with a control problem where the feedback loop has an unknown time delay.

Mathematically, the control problem can be formulated as follows. Let us denote the height measured from the tip as $x(t)$. $x(t)$ is governed by a simple first-order differential equation
\begin{equation}
\dot{x}(t) = -\gamma x(t)+f(\omega t)+u(t),
\label{main}
\end{equation}
where $\gamma>0$ is a stiffness of the cantilever, $\omega$ is a rotation angular velocity, $f(\omega t+T)=f(\omega t)=\sum_{i=0}^{+\infty}f_i \sin(i\omega t+\varphi_i)$ is a $T$-periodic force containing the information on the topography of the surface and the tilting angle, and $u(t)$ represents the control force appears due to the up-and-down lifting mechanism. The parameter $\gamma$ is assumed to be much larger than $\omega$. In fact, in the experiment~\cite{Ulcinas2017,VAITEKONIS2025117552}, $\omega$ is set to approximately 50 Hz while $\gamma$ is of the order of 100 kHz. Therefore, $\omega/\gamma<1$ is justifiable. Note that in Eq.~(\ref{main}), we omitted any factor next to the term $u(t)$, meaning that the external force comes with a sensitivity factor equal to one. It is motivated by the fact that the variable $x(t)$ can always be re-scaled in such a way that Eq.~(\ref{main}) is well justified. Note that our tip-sample interaction model~(\ref{main}) is very simple, yet it is enough to capture the main properties of the tip dynamics. For further extensions of the tip-sample interaction model, see Refs.~\cite{4294022,6913559}.

For the control-free case ($u(t)=0$) solution of Eq.~(\ref{main}) can be written as a power series expansion of the ratio $\omega/\gamma$:
\begin{equation}
x(t)=x_0(t)+x_1(t)+\mathcal{O}\left((\omega/\gamma)^{2}\right).
\label{expan}
\end{equation}
Substitution of the last expression to Eq.~(\ref{main}) yields $x_0(t)=0$ and $x_1(t)=f(\omega t)/\gamma$ meaning that $x(t)\approx f(\omega t)/\gamma$, and therefore $x(t)$ resembles a shape of the roughness of the surface. Restriction to the zero and the first order terms in expansion~(\ref{expan}) is equivalent to a simplification of Eq.~(\ref{main}) up to an equation
\begin{equation}
0 = -\gamma x(t)+f(\omega t)+u(t).
\label{main_s}
\end{equation}
In other words, the plant transfer function $P_0(s)=1/(s+\gamma)$ simplified to $P_0(s)=1/\gamma$. Further in the paper, in all analytical derivation, we will always use simplified Eq.~(\ref{main_s}) instead of full Eq.~(\ref{main}), except in numerical calculations to show that the controller developed for~(\ref{main_s}) works well for the systems~(\ref{main}) if the condition $\omega/\gamma<1$ holds.

\begin{figure}[h!]
\centering\includegraphics[width=0.95\columnwidth]{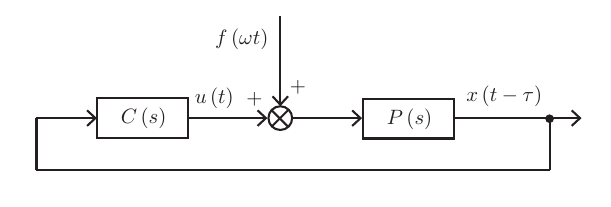}
\caption{\label{fig0} The block scheme of the plant $P(s)=P_0(s)\mathrm{e}^{-\tau s}$ disturbed by the external periodic force $f(\omega t)$ and controlled by the feedback controller $C(s)$.}
\end{figure}

Our goal is to design a linear time-invariant controller $C(s)$ (see Figure~\ref{fig0}) accepting as an input the delayed signal $x(t-\tau)$ and producing the output $u(t)$ such that Eq.~(\ref{main_s}) yields the solution
\begin{equation}
x(t)=\frac{f_0}{\gamma}+\sum_{i=2}^{+\infty}\frac{f_i}{\gamma} \sin(i\omega t+\varphi_i),
\label{sol}
\end{equation}
i.e., $x(t)$ expanded into a Fourier series reproduces the roughness of the surface without the first Fourier harmonic. The time delay $\tau$ is assumed to be unknown, yet not larger than the period $T=2\pi/\omega$, so that $\tau/T<1$.

\section{\label{subsec:cont}Design of the controller}

As a central idea of our control algorithm, we will use the controller developed in Refs.~\cite{Olyaei2015,AzimiOlyaei2018}. The controller contains a system of harmonic oscillators coupled with the input signal. The dynamical variables of the oscillators are used to design the output signal devoted to stabilize an unknown unstable periodic orbit in a dynamical system. Here, our goal is different. Instead of stabilization, we want the elimination of the first Fourier harmonic. Therefore, we will modify the controller~\cite{Olyaei2015,AzimiOlyaei2018} to fulfill our goal.

\subsection{\label{subsec:one_har}Restriction to the case of one harmonic in the force $f(\omega t)$}

Firstly, one can solve a simplified task. In particular, let us say that the force $f(\omega t)$ contains only the first harmonic, i.e., we have $f(\omega t)=f_1 \sin(\omega t)$. Here and further in the text, we set the initial phase $\varphi_1=0$ without loss of generality. For such a case, one can use only one harmonic oscillator coupled with the delayed system output $x(t-\tau)$:
\begin{subequations}
\label{one_harm}
\begin{align}
\dot{a}_1(t) &= -\omega b_1(t)+\alpha_1 x(t-\tau), \label{one_harm_1} \\
\dot{b}_1(t) &= \omega a_1(t)+\beta_1 x(t-\tau). \label{one_harm_2}
\end{align}
\end{subequations}
Here $a_1(t)$ and $b_1(t)$ are the real-valued dynamical variables of the harmonic oscillator while $\alpha_1$ and $\beta_1$ are the coupling constants. The output of the controller
\begin{equation}
u(t)=K a_1(t),
\label{out}
\end{equation}
where $K$ stands for the control gain. The closed loop system~(\ref{main_s}), (\ref{one_harm}) and (\ref{out}) possesses our desired solution
\begin{subequations}
\label{sol_one}
\begin{align}
x(t) &= 0,\label{sol_one_1} \\
a_1(t) &= -\frac{f_1}{K} \sin(\omega t), \label{sol_one_2} \\
b_1(t) &= \frac{f_1}{K} \cos(\omega t). \label{sol_one_3}
\end{align}
\end{subequations}
Small deviations $\delta x(t)$, $\delta a_1(t)$ and $\delta b_1(t)$ from the desired solution~(\ref{sol_one}) satisfy following linear time-invariant system of equations
\begin{subequations}
\label{pert}
\begin{align}
\delta x(t) &= \frac{K}{\gamma} \delta a_1(t),\label{pert_1} \\
\delta\dot{a}_1(t) &= -\omega \delta b_1(t)+\alpha_1 \delta x(t-\tau), \label{pert_2} \\
\delta \dot{b}_1(t) &= \omega \delta a_1(t)+\beta_1 \delta x(t-\tau). \label{pert_3}
\end{align}
\end{subequations}
Using the Laplace transform and the formalism of the transfer function, one can formulate the stability of the closed-loop system~(\ref{pert}). We will use standard uppercase letter notation for the Laplace transformed signals. In particular, if we have a signal $y(t)$ in the time domain, then the Laplace transformed signal in the frequency domain denoted as $Y(s)=\mathcal{L}(y)=\int_0^{+\infty} y(t)\exp(-st) \mathrm{d}t$.

Simple Eq.~(\ref{pert_1}) defines the dynamics of a plant (system to be controlled). The plant as an input signal accepts $\delta u(t)=K \delta a_1(t)$ and produces an output signal $\delta x(t-\tau)$ which is delayed plant variable $\delta x(t)$. Therefore, the plant transfer function
\begin{equation}
P(s)=\frac{\delta X(s) \mathrm{e}^{-\tau s}}{\delta U(s)}=\frac{\delta X(s) \mathrm{e}^{-\tau s}}{K \delta A_1(s)}=\frac{\mathrm{e}^{-\tau s}}{\gamma}.
\label{plant_trf}
\end{equation}
The controller described by Eqs.~(\ref{pert_2}), (\ref{pert_3}) accepts $\delta x(t-\tau)$ as an input signal and produce $\delta u(t)=K \delta a_1(t)$ as an output signal. The transfer function of the controller is
\begin{equation}
C(s)=K\frac{\delta A_1(s)}{\delta X(s) \mathrm{e}^{-\tau s}}=K \frac{s\alpha_1-\omega \beta_1}{s^2+\omega^2}.
\label{cont_trf}
\end{equation}
It is important to mention that the controller of full variables~(\ref{one_harm}) and the controller for small deviations~(\ref{pert_2}), (\ref{pert_3}) coincides: both systems are linear time-invariant therefore has the same transfer function, meaning that the same Eq.~(\ref{cont_trf}) would be derived for the controller's~(\ref{one_harm}) transfer function $C(s)=K A_1(s)/[X(s)\mathrm{e}^{-\tau s}]$. In contrast, the plant of full variables~(\ref{main_s}) is not time-invariant due to the term $f(\omega t)$. Therefore, we derived Eq.~(\ref{pert_1}), which is time-invariant.

By having the plant transfer function~(\ref{plant_trf}) and the controller transfer function~(\ref{cont_trf}), one can write the transfer function of the whole closed-loop system~(\ref{pert})
\begin{equation}
\begin{aligned}
G(s) &= \frac{C(s)P(s)}{1-C(s)P(s)} \\
&= \frac{K}{\gamma} \frac{ \frac{s}{\omega} \bar{\alpha}_1- \bar{\beta}_1}{\exp\left[2\pi \frac{\tau}{T}  \frac{s}{\omega}\right] \left(\left( \frac{s}{\omega} \right)^2+1 \right)-\frac{K}{\gamma} \left(\frac{s}{\omega} \bar{\alpha}_1-\bar{\beta}_1 \right)},
\label{close_trf}
\end{aligned}
\end{equation}
where we introduced normalized coupling constants $\bar{\alpha}_1=\alpha_1/\omega$, $\bar{\beta}_1=\beta_1/\omega$. The stability of the system~(\ref{close_trf}) is determined by the real parts of poles; namely, the system is stable if and only if all poles have negative real parts. Let us analyze only positive $K$ values, since from Eq.~(\ref{cont_trf}) one can see that the negative $K$ reproduces the same $C(s)$ if we flip the signs for both parameters $\alpha_1$ and $\beta_1$. In the case $\tau=0$, the equation for the poles is just a quadratic equation,
\begin{equation}
\left(\frac{s}{\omega}\right)^2+1-\frac{K}{\gamma} \left(\frac{s}{\omega} \bar{\alpha}_1-\bar{\beta}_1 \right)=0,
\label{poles}
\end{equation}
yielding the necessary and sufficient stability conditions,
\begin{equation}
\bar{\alpha}_1<0 \quad \mathrm{and} \quad \frac{K}{\gamma} \bar{\beta}_1 > -1.
\label{stab_cond}
\end{equation}
\begin{figure*}
\includegraphics[width=0.95\textwidth]{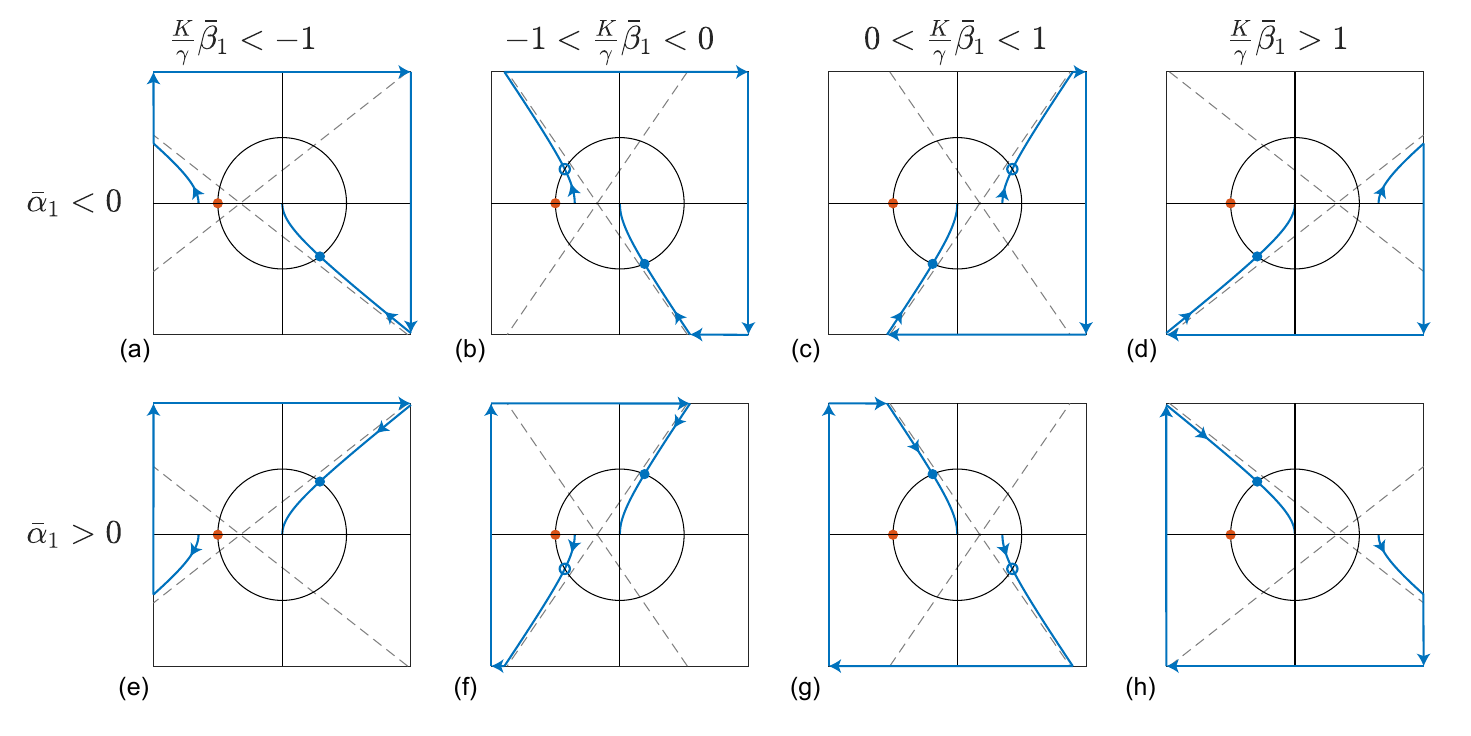}
\caption{\label{figNyq} The Nyquist plots of the open-loop system (\ref{open_trf}) for 8 topologically different cases when the frequency $\Omega$ increases from 0 to $+\infty$. The part of the Nyquist plot for $\Omega=-\infty$ to $\Omega=0$ can be obtained by reflecting the drawn part with respect to the $x$-axis. Each row corresponds to $\bar{\alpha}_1<0$ or $\bar{\alpha}_1>0$ while each column corresponds to four different cases, namely, $(K/\gamma)\bar{\beta}_1<-1$, $-1<(K/\gamma)\bar{\beta}_1<0$, $0<(K/\gamma)\bar{\beta}_1<1$ and $(K/\gamma)\bar{\beta}_1>1$. Asymptotes of the hyperbola are drawn by dashed grey lines. The red closed circle represents the point $(-1,0)$, while the black line depicts the unit circle. The critical frequencies $\Omega_{1}$ and $\Omega_{2}$ when the Nyquist curve goes outside and inside into the unit circle, $|G_0(\mathrm{i}\Omega_{1,2})|=1$, depicted by an open and closed circle, respectively.}
\end{figure*}
To analyze the system's stability for a non-zero delay, one can use the Nyquist stability criterion of the delay-free system and then obtain critical delay values $\tau$ when the system loses its stability~\cite{tsyp46}. The delay-free open-loop transfer function
\begin{equation}
G_0(\mathrm{i}\Omega)=-C(\mathrm{i}\Omega)P(\mathrm{i}\Omega)= \frac{K}{\gamma} \frac{\bar{\beta}_1-\mathrm{i}\frac{\Omega}{\omega} \bar{\alpha}_1}{1-\left(\frac{\Omega}{\omega}\right)^2}.
\label{open_trf}
\end{equation}
The Nyquist plot of~(\ref{open_trf}) is a hyperbola with a shifted origin and re-scaled $x$ and $y$ axes. Indeed, for the new variables
\begin{subequations}
\label{hyper_var}
\begin{align}
x_0(\Omega) &= \frac{\Re\left[ G_0(\mathrm{i}\Omega) \right]}{\bar{\beta}_1} - \frac{1}{2}\frac{K}{\gamma} = \frac{1}{2}\frac{K}{\gamma}\frac{1+\left(\frac{\Omega}{\omega}\right)^2}{1-\left(\frac{\Omega}{\omega}\right)^2},\label{hyper_var_1} \\
y_0(\Omega) &= -\frac{\Im\left[ G_0(\mathrm{i}\Omega) \right]}{\bar{\alpha}_1}=\frac{K}{\gamma}\frac{\frac{\Omega}{\omega}}{1-\left(\frac{\Omega}{\omega}\right)^2}, \label{hyper_var_2}
\end{align}
\end{subequations}
we obtain a hyperbolic equation
\begin{equation}
x_0^2(\Omega)-y_0^2(\Omega)=\left(\frac{K}{2\gamma}\right)^2.
\label{hyper}
\end{equation}
The open-loop system, $G_0(\mathrm{i}\Omega)$, has two poles on the imaginary axis at $\Omega/\omega=\pm 1$. Therefore, for the part of the Nyquist contour from $\Omega=0$ to $\Omega=+\infty$, we modify the contour in the small vicinity of the point $\Omega/\omega=1$ by going around the pole, leaving the pole outside of the contour. Such a manipulation gives a clockwise rotation of the complex phase by $\pi$ in the Nyquist plot.

From Eq.~(\ref{open_trf}), we can see that we should analyze eight topologically different cases of the Nyquist plots. Figure~\ref{figNyq} shows schematic representations of the Nyquist plots for all 8 cases. As we can see, the delay-free system is stable for the cases Figure~\ref{figNyq}~(b),~(c),~(d), and unstable otherwise, which coincides with~(\ref{stab_cond}). The time delay added to the system gives the same Nyquist plot, only the phase is rotated clockwise. The critical frequencies $\Omega_{1,2}$, when $|G_0(\mathrm{i}\Omega_{1,2})|=1$, can be obtained from a quadratic equation
\begin{equation}
\left(\frac{\Omega}{\omega}\right)^4-\left[2+ \left(\frac{K}{\gamma}\right)^2 \bar{\alpha}_1^2 \right]\left(\frac{\Omega}{\omega}\right)^2+1- \left(\frac{K}{\gamma}\right)^2 \bar{\beta}_1^2=0
\label{crit_freq_quad}
\end{equation}
yielding two solutions
\begin{widetext}
\begin{subequations}
\label{crit_freq}
\begin{align}
\frac{\Omega_1}{\omega} &= \sqrt{1+\left(\frac{K}{\gamma}\right)^2 \left(\frac{\bar{\alpha}_1}{\sqrt{2}} \right)^2 -\left(\frac{K}{\gamma}\right)\sqrt{\left(\frac{K}{\gamma}\right)^2 \left(\frac{\bar{\alpha}_1}{\sqrt{2}} \right)^4 + (\bar{\alpha}_1^2+\bar{\beta}_1^2)}} ,\label{crit_freq_1} \\
\frac{\Omega_2}{\omega} &= \sqrt{1+\left(\frac{K}{\gamma}\right)^2 \left(\frac{\bar{\alpha}_1}{\sqrt{2}} \right)^2 +\left(\frac{K}{\gamma}\right)\sqrt{\left(\frac{K}{\gamma}\right)^2 \left(\frac{\bar{\alpha}_1}{\sqrt{2}} \right)^4 + (\bar{\alpha}_1^2+\bar{\beta}_1^2)}}. \label{crit_freq_2}
\end{align}
\end{subequations}
\end{widetext}
From (\ref{crit_freq_1}), it can be shown that the solution $\Omega_1$ exists only when the starting point of the Nyquist plot is inside the unit circle, $|G_0(0)| \leq 1$, i.e. $\left( K/\gamma \right) |\bar{\beta}_1| \leq 1$. Such a situation realizes in Figure~\ref{figNyq}~(b),~(c),~(f),~(g). In contrast, the solution $\Omega_2$ always exists. Now, we can look at when the time delay changes the stability. The cases Figure~\ref{figNyq}~(a),~(e),~(h) remain unstable for any delay $\tau$. The case Figure~\ref{figNyq}~(b) looses the stability at
\begin{equation}
\cos\left( 2\pi\frac{\Omega_2}{\omega} \frac{\tau}{T} \right) \sqrt{1+\left(\frac{\Omega_2}{\omega} \frac{\bar{\alpha}_1}{\bar{\beta}_1}\right)^2}=-1,
\label{stab_bord1}
\end{equation}
while the cases Figure~\ref{figNyq}~(c),~(d) loose the stability at
\begin{equation}
\cos\left( 2\pi\frac{\Omega_2}{\omega} \frac{\tau}{T} \right) \sqrt{1+\left(\frac{\Omega_2}{\omega} \frac{\bar{\alpha}_1}{\bar{\beta}_1}\right)^2}=1.
\label{stab_bord2}
\end{equation}
The interesting situation happens with the case Figure~\ref{figNyq}~(f): while the delay-free system is unstable, it becomes stable once the delay increases. Such effect is known as delay-induced stability~\cite{Atay2010}, one of the first times observed in Ref.~\cite{4793475} where a similar closed-loop system is analyzed. The stability border for the case Figure~\ref{figNyq}~(f) is
\begin{equation}
\cos\left( 2\pi\frac{\Omega_1}{\omega} \frac{\tau}{T} \right) \sqrt{1+\left(\frac{\Omega_1}{\omega} \frac{\bar{\alpha}_1}{\bar{\beta}_1}\right)^2}=1.
\label{stab_bord3}
\end{equation}
The case Figure~\ref{figNyq}~(g) also can gain the stability at
\begin{equation}
\cos\left( 2\pi\frac{\Omega_1}{\omega} \frac{\tau}{T} \right) \sqrt{1+\left(\frac{\Omega_1}{\omega} \frac{\bar{\alpha}_1}{\bar{\beta}_1}\right)^2}=-1.
\label{stab_bord4}
\end{equation}
However, such a situation is not in our field of interest, since it appears at high enough delay values.
\begin{figure}[h!]
\centering\includegraphics[width=0.85\columnwidth]{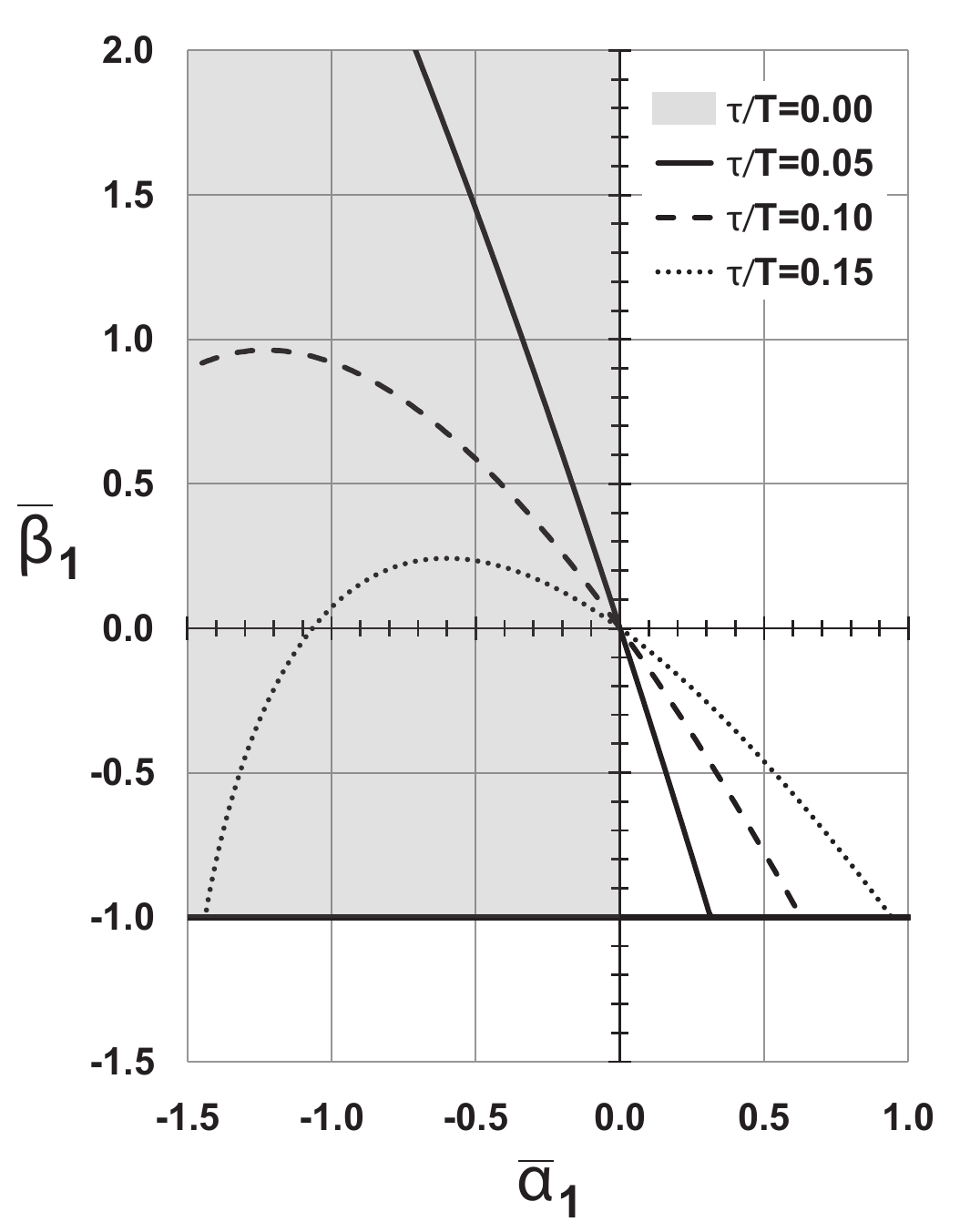}
\caption{\label{fig1a} The closed-loop systems~(\ref{close_trf}) stability region calculated numerically from Eqs.~(\ref{stab_bord1}),~(\ref{stab_bord2}),~(\ref{stab_bord3}) with different values of the ratio $\tau/T$. The parameter $K/\gamma$ is set to $K/\gamma=1$.}
\end{figure}

In Figure~\ref{fig1a}, we plot stability regions for several different values of the ratio $\tau/T$ by numerically solving Eqs.~(\ref{stab_bord1}),~(\ref{stab_bord2}),~(\ref{stab_bord3}). One can see that the horizontal stability border does not change by increasing $\tau/T$ while the vertical stability border modifies to a parabola-like curve. In order to obtain it, for the region $\bar{\beta}_1>0$ we use Eq.~(\ref{stab_bord2}), while for the region $\bar{\alpha}_1<0$ and $-1<\bar{\beta}_1<0$ we use Eq.~(\ref{stab_bord1}). The triangular shape stability region $\bar{\alpha}_1>0$ and $-1<\bar{\beta}_1<0$, where delay induced stability appears, is found from Eq.~(\ref{stab_bord3}).

\begin{figure}[h!]
\centering\includegraphics[width=0.95\columnwidth]{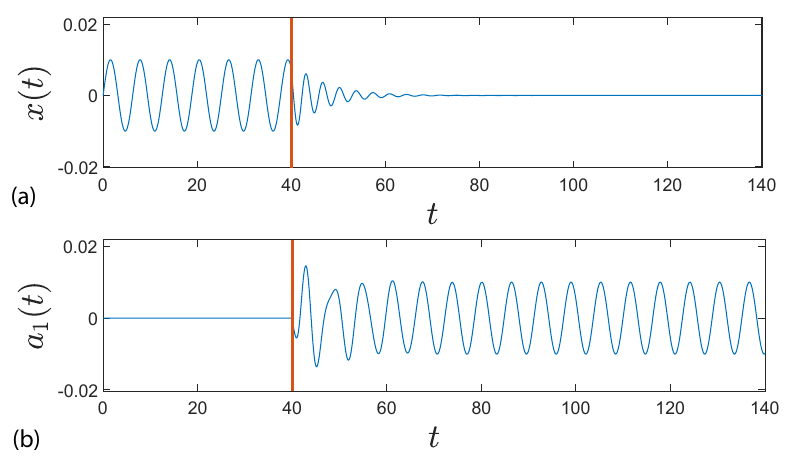}
\caption{\label{fig2} The results of the numerical simulation of closed-loop system~(\ref{main}), (\ref{one_harm}) and (\ref{out}). Before the time moment $t=40$ the system was in a control-free regime ($K=0$). (a) -- the dynamics of the plant variable, (b) -- the same for the first controller variable $a_1(t)$. The parameters are following: $K/\gamma=1$, $\gamma=200$, $f_1=1$, $\omega=1$, $\alpha_1=-1$, $\beta_1=-0.5$ and the time delay $\tau/T=0.15$.}
\end{figure}

To demonstrate the successful work of our controller, in Figure~\ref{fig2}, we depicted a numerical simulation of the plant equation~(\ref{main}) with $f(\omega t)=f_1\sin(\omega t)$ and feedback loop defined by Eqs.~(\ref{one_harm}) and (\ref{out}). As one can see, the controller achieves stabilization for non-zero delay $\tau/T=0.15$ and the parameters $(\alpha_1,\beta_1)=(-1,-0.5)$. Yet, a further increase in the time delay leads to a loss of stability.

\subsection{\label{subsec:all_har}The case of full set of harmonics}

This time we assume that the force $f(\omega t)$ contains a set of $N$ Fourier harmonics
\begin{equation}
f(\omega t)=\sum_{i=0}^{N} f_i \sin\left( i \omega t+\varphi_i \right).
\label{full_f}
\end{equation}
In particular, $N$ can be equal to $+\infty$, meaning that all harmonics contribute to the force $f$. Similarly to previous Subsection~\ref{subsec:one_har}, without loss of generality, we can assume that $\varphi_1=0$. Since we have $N$ Fourier harmonics, our controller should contain a set of $N$ harmonic oscillators coupled to the input signal $x(t-\tau)$. For such a case, Eqs.~(\ref{one_harm}) can be generalized as follows:
\begin{subequations}
\label{all_harm}
\begin{align}
\dot{a}_0(t) &= \alpha_0 \left[ x(t-\tau) - \sum_{\substack{i=0 \\ i \neq 1}}^{N} a_i(t) \right], \label{all_harm_1} \\
\dot{a}_j(t) &= -j \omega b_j(t)+\alpha_j \left[ x(t-\tau) - \sum_{\substack{i=0 \\ i \neq 1}}^{N} a_i(t) \right], \label{all_harm_2} \\
\dot{b}_j(t) &= j \omega a_j(t)+\beta_j \left[ x(t-\tau) - \sum_{\substack{i=0 \\ i \neq 1}}^{N} a_i(t) \right], \label{all_harm_3}
\end{align}
\end{subequations}
where $j=1,2,\ldots ,N$. Here $\alpha_j$ and $\beta_j$ are coupling constants to be determined below. The controller output has the same form~(\ref{out}) as in the one harmonic case.

The closed loop system~(\ref{main_s}), (\ref{all_harm}) and (\ref{out}) possesses the solution of $x(t)$ without the first harmonic
\begin{subequations}
\label{sol_all}
\begin{align}
x(t) &= \frac{1}{\gamma} \sum_{i=0 \atop i \neq 1}^{N} f_i \sin\left( i \omega t +\varphi_i \right) , \label{sol_all_1} \\
a_0(t) &= \frac{f_0}{\gamma}, \label{sol_all_2} \\
a_1(t) &= -\frac{f_1}{K}\sin( \omega t), \label{sol_all_3} \\
b_1(t) &= \frac{f_1}{K}\cos( \omega t), \label{sol_all_4} \\
a_j(t) &= \frac{f_j}{\gamma}\sin\left( j \omega (t-\tau) +\varphi_j \right), \label{sol_all_5} \\
b_j(t) &= -\frac{f_j}{\gamma}\cos\left( j \omega (t-\tau) +\varphi_j \right), \label{sol_all_6}
\end{align}
\end{subequations}
with $j=2,3,\ldots,N$. While the controller equations~(\ref{all_harm}) is the linear time-invariant system, the plant~(\ref{main_s}) is linear but not a time-invariant system. To treat the system using transfer function formalism, one should have a linear time-invariant system. Therefore, one can write the equations for the small deviation from the target solution~(\ref{sol_all}):
\begin{subequations}
\label{pert_all}
\begin{align}
\delta x(t) &= \frac{K}{\gamma} \delta a_1(t), \label{pert_all_1} \\
\delta \dot{a}_0(t) &= \alpha_0 \left[ \delta x(t-\tau) - \sum_{i=0 \atop i \neq 1}^{N} \delta a_i(t) \right], \label{pert_all_2} \\
\delta \dot{a}_j(t) &= -j \omega \delta b_j(t)+\alpha_j \left[ \delta x(t-\tau) - \sum_{i=0 \atop i \neq 1}^{N} \delta a_i(t) \right], \label{pert_all_3} \\
\delta \dot{b}_j(t) &= j \omega \delta a_j(t)+\beta_j \left[ \delta x(t-\tau) - \sum_{i=0 \atop i \neq 1}^{N} \delta a_i(t) \right], \label{pert_all_4}
\end{align}
\end{subequations}
with $j=1,2,\ldots,N$. The transfer function of the plant system~(\ref{pert_all_1}) has the same simple form~(\ref{plant_trf}) as in previous Subsection~\ref{subsec:one_har}. However, the controller transfer function $C(s)=\delta U(s)/[\delta X(s) \mathrm{e}^{-\tau s}]$ is much more complicated. In the Appendix~\ref{app:trans}, we provide a derivation of the controller's transfer function:
\begin{equation}
\begin{aligned}
C(s) =& K \left(\alpha_1 s-\beta_1 \omega \right) s \prod_{j=2}^N \left( s^2+ j^2 \omega^2 \right) \\
&\times \left\lbrace (s+\alpha_0) \prod_{j=1}^N \left( s^2+ j^2 \omega^2 \right) \right. \\
&+ \left. s \sum_{k=2}^{N} \left(\alpha_k s- k \beta_k \omega \right) \prod_{j=1 \atop j \neq k}^N \left( s^2+ j^2 \omega^2 \right) \right\rbrace^{-1}.
\label{cont_trans_f}
\end{aligned}
\end{equation}
The zeros of the transfer function are $s=0$, $s=\beta_1 \omega/\alpha_1$, $s=\pm \mathrm{i}k\omega$ for $k=2,3,\ldots,N$, while the expression in the curly brackets determines the poles. The closed-loop system transfer function $G(s)$ can be obtained by the same formula~(\ref{close_trf}). The stability of the solution~(\ref{sol_all}) is determined by the poles of the function $G(s)$ involving both the zeros and the poles of the controller's transfer function $C(s)$. To be more precise, the poles of $G(s)$ are solutions for the equation
\begin{equation}
\begin{aligned}
&\gamma(s+\alpha_0) \prod_{j=1}^N \left( s^2+ j^2 \omega^2 \right) \\
&+ \gamma s \sum_{k=2}^{N} \left(\alpha_k s- k \beta_k \omega \right) \prod_{j=1 \atop j \neq k}^N \left( s^2+ j^2 \omega^2 \right)\\
&- K \mathrm{e}^{-\tau s} s \left(\alpha_1 s-\beta_1 \omega \right) \prod_{j=2}^N \left( s^2+ j^2 \omega^2 \right)=0 .
\label{poles1}
\end{aligned}
\end{equation}
Such an equation has many parameters. To simplify it, we set some particular form of the coupling constants $\alpha_k$ and $\beta_k$: 
\begin{equation}
\begin{aligned}
&\alpha_0 = \alpha, \; \alpha_1=-2\alpha\frac{\gamma}{K}, \; \beta_1 = 2\beta\frac{\gamma}{K}, \\
&\alpha_k=2\alpha, \; \beta_k = -2\beta/k \quad \mathrm{for} \; k=2 \ldots N.
\label{cons_form}
\end{aligned}
\end{equation}
Therefore, we have only two real index-less variables $\alpha$ and $\beta$. The motivation behind such a form is that it significantly simplifies stability analysis. Now Eq.~(\ref{poles1}) reduces to
\begin{equation}
\begin{aligned}
&(s+\alpha) \prod_{j=1}^N \left( s^2+ j^2 \omega^2 \right) \\
&+ 2s \left(\alpha s + \beta \omega \right) \sum_{k=1}^{N} \mathrm{e}^{-\tau s \delta_{1,k}}  \prod_{j=1 \atop j \neq k}^N \left( s^2+ j^2 \omega^2 \right)=0,
\label{poles2}
\end{aligned}
\end{equation}
where $\delta_{1,k}$ is non-zero only for $k=1$. Similar to Eq.~(\ref{poles}), to analytically find the poles of the closed-loop transfer function $G(s)$, we consider the simplest case $\tau=0$. Now Eq.~(\ref{poles2}) reads
\begin{equation}
(s+\alpha) q(s)+ \left(\alpha s + \beta \omega \right) \frac{\mathrm{d}q(s)}{\mathrm{d}s}=0,
\label{poles3}
\end{equation}
where the function $q(s)$ is defined as
\begin{equation}
q(s)=\prod_{j=1}^N \left( s^2+ j^2 \omega^2 \right).
\label{q}
\end{equation}
In the Appendix~\ref{app:stab}, we show that in the limit $N\rightarrow + \infty$, the necessary and sufficient condition for the roots of Eq.~(\ref{poles3}) to have negative real parts is
\begin{equation}
\alpha>0 \quad \mathrm{and} \quad \frac{\beta}{\omega} > -\frac{1}{4}.
\label{stab_cond1}
\end{equation}
In contrast to the previous stability condition~(\ref{stab_cond}), here the stability border goes along $\beta/\omega=-1/4$ while in the one harmonic case~(\ref{stab_cond}) it goes along $\beta/\omega=-1/2$ (note that the relation of the index-less coupling constants $\alpha$ and $\beta$ with $\alpha_1$ and $\beta_1$ is defined by Eq.~(\ref{cons_form})).
\begin{figure}[h!]
\centering\includegraphics[width=0.85\columnwidth]{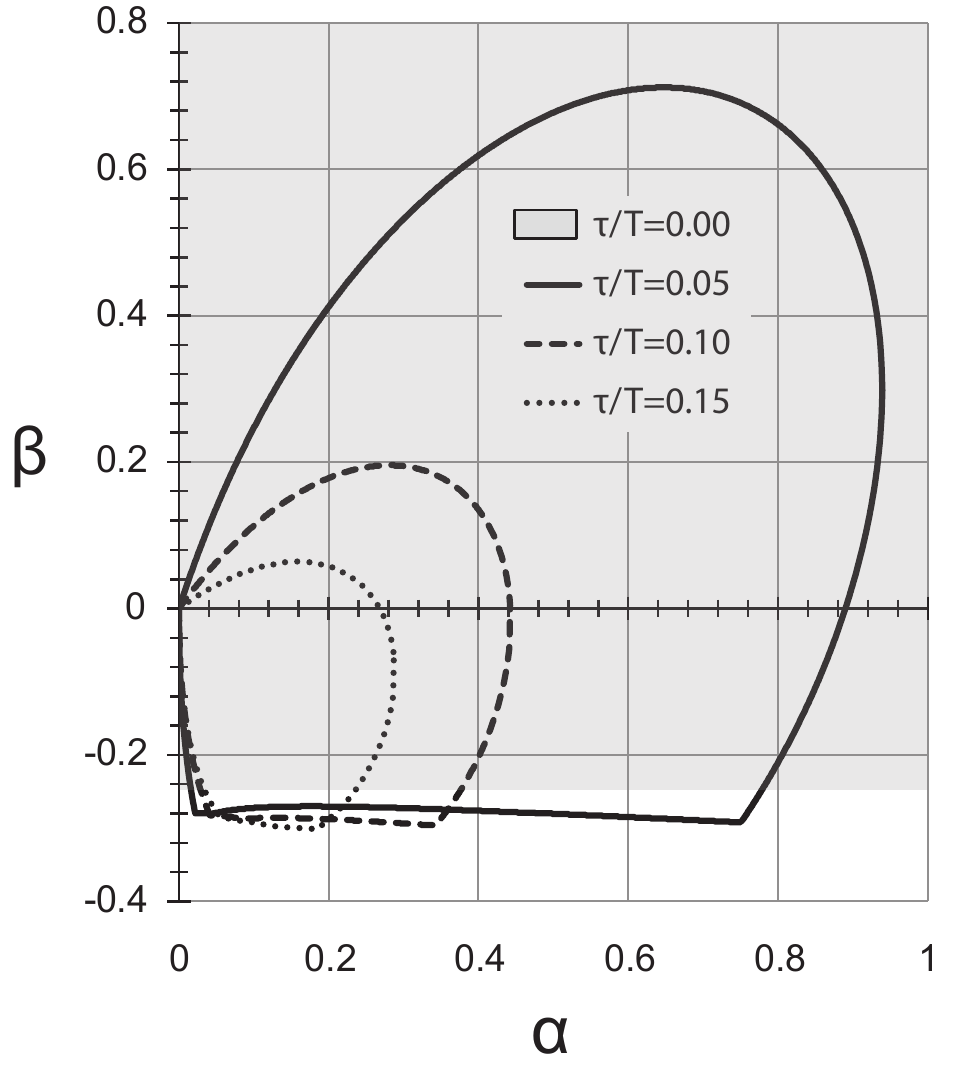}
\caption{\label{fig3} The stability regions for the roots of Eq.~(\ref{poles2}) calculated numerically in the limit $N\rightarrow + \infty$. The frequency $\omega$ is set to be equal to 1.}
\end{figure}

The Nyquist stability criterion is problematic for the non-zero time delay because of $2N+1$ zeros on the imaginary axes for the open-loop transfer function. Therefore, we numerically calculated the stability region. We analyzed Eq.~(\ref{poles2}) in the limit $N\rightarrow +\infty$ and calculated events when the pair (or several pairs) of complex conjugate roots crossed the imaginary axis, denoting loss of stability. In Figure~\ref{fig3}, we plot borders of the stability regions for several values of the ratio $\tau/T$. As one can see, the stability region significantly shrinks once we increase the time delay. Yet we point out that there are ``universal'' values $(\alpha/\omega,\beta/\omega)=(0.1, -0.1)$ where the system remains stable at least up to $\tau/T=0.15$, making them a good starting point in an actual experimental situation.

\begin{figure}[h!]
\centering\includegraphics[width=0.95\columnwidth]{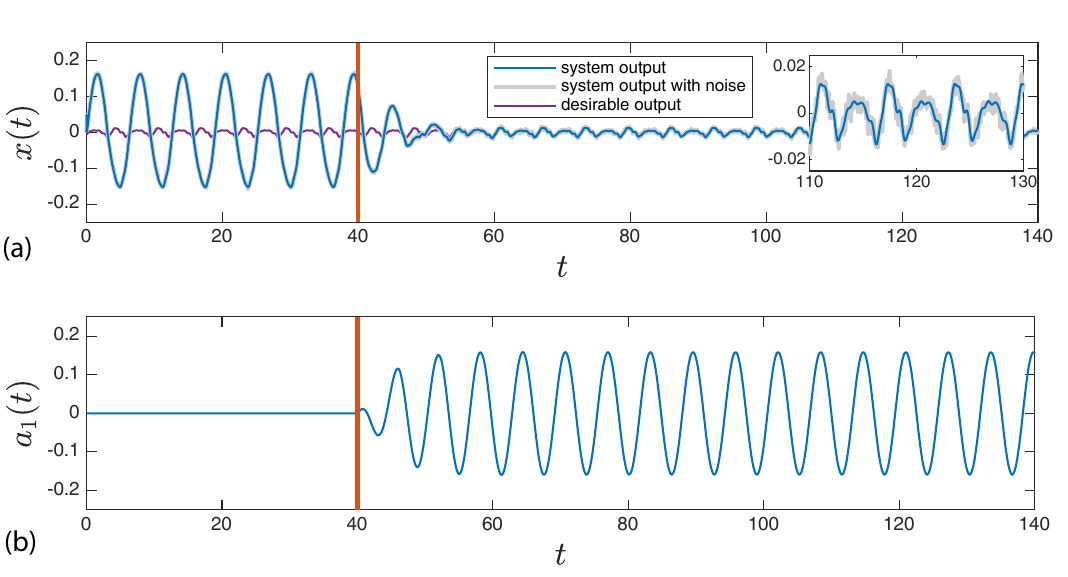}
\caption{\label{fig4} The results of the numerical simulation of closed-loop system~(\ref{main}), (\ref{full_f}), (\ref{all_harm}) and (\ref{out}) with the finite number of the harmonics, $N=10$. Before the time moment $t=40$, the system is in a control-free regime ($K=0$). (a) -- the dynamics of the plant variable (thin blue curve), the dynamic of the plant variable when the external force is supplemented with a noise (thick grey line) and the external force without the first harmonic $f_0/\gamma+\sum_{i=2}^N f_i/\gamma \sin(i\omega t+\varphi_i)$ (purple line), the inset graphics shows that a noisy dynamics follows desirable output. (b) -- the first controller variable $a_1(t)$ for the dynamics without noise. The parameters are the following: $K/\gamma=1$, $\gamma=100$, $\omega=1$, $\alpha=0.1$, $\beta=-0.1$, $\tau/T=0.15$, the amplitudes $f_i$ and the phases $\varphi_i$ generated randomly except $f_1$, which is artificially increased to better express the successful work of the controller. The phases $\varphi_i$ are generated from a uniform distribution of the interval $[0,2\pi]$, while the amplitudes $f_i$ are generated from the uniform distribution of the interval $[0,1]$ and additionally divided by the harmonic number $i$. The noise for the external force is generated as a Gaussian uncorrelated stochastic force with a zero mean and a unit standard deviation.}
\end{figure}

To demonstrate the successful application of the controller, we numerically integrate the plant~(\ref{main}) with the external force $f(\omega t)$ defined by Eq.~(\ref{full_f}) and the controller defined by Eqs.~(\ref{all_harm}), (\ref{out}). In Figure~\ref{fig4}~(a), we plotted the plant variable $x(t)$ and compared it with the external force without the first Fourier harmonic $[f(\omega t)-f_1 \sin(\omega t)]/\gamma$. To show robustness against noise, in Figure~\ref{fig4}~(a) we plotted the plant variable when a stochastic term supplements the external force. In the case of successful first harmonic elimination, according to Eq.~(\ref{sol_all_1}), the plant variable should settle to the external filtered force. This is precisely what we observe in Figure~\ref{fig4}~(a) after $t=40$ and the transient time. The time delay is set to $\tau/T=0.15$. Further enlargement of the time delay will destabilize the closed-loop system.

\section{\label{sec:block_scheme}The block scheme and frequency response of the controller}

\begin{figure}[h!]
\centering\includegraphics[width=0.95\columnwidth]{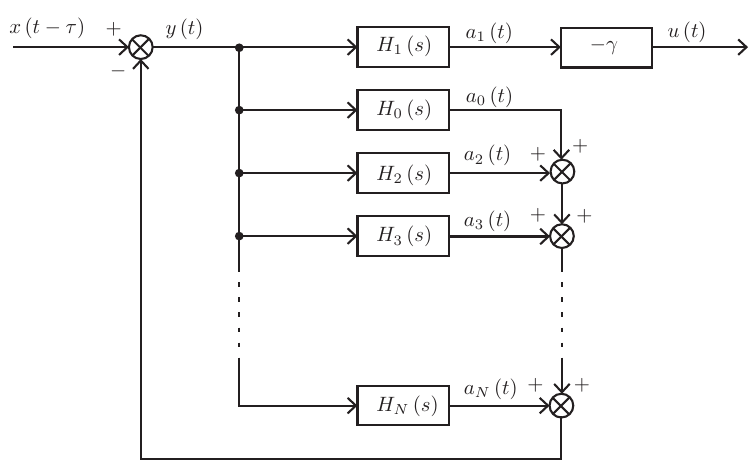}
\caption{\label{fig6} The block scheme representing the controller~(\ref{cont_trans_f2}) containing the set of the harmonic oscillators $H_j(s)$ defined in Eqs~(\ref{hj}), (\ref{h0}).}
\end{figure}

The controller described by Eqs.~(\ref{all_harm}) is the linear time-invariant controller with the transfer function~(\ref{cont_trans_f}). Taking in to account the form of coupling constants~(\ref{cons_form}) we obtain the transfer function
\begin{equation}
C(s) = (-\gamma) 2\left( \frac{\alpha s + \beta \omega}{s^2+\omega^2} \right)  \left\lbrace 1+ \frac{\alpha}{s} + 2\sum_{j=2}^{N} \frac{\alpha s+ \beta \omega}{s^2+ j^2\omega^2}  \right\rbrace^{-1} .
\label{cont_trans_f2}
\end{equation}
Such transfer function can be represented by the block scheme depicted in Figure~\ref{fig6}, where
\begin{equation}
H_j(s) =  2\frac{\alpha s +\beta \omega}{s^2+j^2 \omega^2}
\label{hj}
\end{equation}
for $j=1,2,\ldots,N$ and
\begin{equation}
H_0(s) =  \frac{\alpha}{s}
\label{h0}
\end{equation}
are the transfer functions of the harmonic oscillators. The logical description of the block scheme can be as follows: Let us assume that initial conditions for the oscillators $H_{0,2,3,\ldots,N}(s)$ are perfectly matched with the signal $x(t-\tau)$ such that the signal $y(t)$ possesses only the first harmonic. If so happens, $y(t)$ disturbs the first oscillator $H_1(s)$ until its amplitude and phase produce a correct output $-\gamma a_1(t)$ eliminating the first harmonic in $x(t-\tau)$ and, therefore making $y(t)=0$. Note that the signal $y(t)$ is exactly the term in square brackets of Eq.~(\ref{all_harm}). If, on the other hand, the initial conditions for $H_{0,2,3,\ldots,N}(s)$ are not adjusted with $x(t-\tau)$, then $y(t)$ is non-zero, therefore forcing the oscillators to the correct states.

\begin{figure}[h!]
\centering\includegraphics[width=0.95\columnwidth]{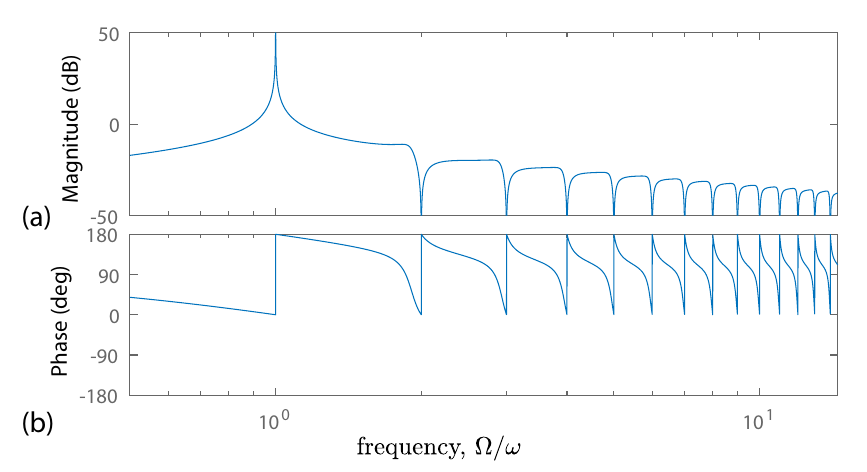}
\caption{\label{fig6a} The Bode plot for the controller (\ref{cont_trans_f2}) with the parameters $\alpha/\omega=0.1$, $\beta/\omega=-0.1$, $\gamma=1$ and $N=16$. (a) represents $|C(s=\mathrm{i}\Omega)|$ in decibels, (b) shows $\mathrm{arg}[C(s=\mathrm{i}\Omega)]$ in degrees.}
\end{figure}

The frequency response $C(\mathrm{i}\Omega)$ of the controller is depicted in Figure~\ref{fig6a}. The gain is infinite at $\Omega=\omega$ and zero at $\Omega = j\omega$ for $j=0,2,\ldots,N$. Basically, the controller is an inverse notch filter at the fundamental frequency and regular notch filters at other integer multiples of the frequency. This is well consistent with the objective of the controller: to reject the first harmonic and not disturb all other harmonics. Indeed, the transfer function from the external disturbance, $f(\omega t)$, to the system's output, $x(t-\tau)$, reads
\begin{equation}
G_{f \rightarrow x} (s) = \frac{X(s)\mathrm{e}^{-\tau S}}{F(s)}=\frac{P(s)}{[1-P(s)C(s)]},
\label{g_gx}
\end{equation}
giving $G_{f \rightarrow x}(s=\mathrm{i}\omega)=0$ (rejection of the first harmonic) and $G_{f \rightarrow x}(s)=P(s)$ (the harmonics remains undistorted) at $s=0,\pm 2\mathrm{i}\omega, \pm 3\mathrm{i}\omega, \ldots \pm N\mathrm{i}\omega$.

\section{\label{sec:infinite}Controller's equation for the infinite number of the harmonics}

Similar to the Refs.~\cite{Escobar2005,Lu2010,PyrHarmOsc2015}, we can rewrite our transfer function~(\ref{cont_trans_f2}) in the limit when the number of harmonic oscillators goes to infinity. Then, we can obtain a controller scheme in the time domain using the inverse Laplace transform. Such a scheme can be written as delay differential algebraic equations instead of having an infinite number of ordinary differential equations, as in Eqs.~(\ref{all_harm}). Therefore, the number of dynamical variables remains finite, yet the infinite dimension is encoded via delayed terms.

Using notation~(\ref{q}), the transfer function reads
\begin{equation}
\begin{aligned}
C(s) =& -2\gamma \left( \alpha s + \beta \omega \right) s q(s) \\
&\times \left\lbrace (s+\alpha)\left( s^2+\omega^2 \right) q(s) - 2 \left( \alpha s + \beta \omega \right) s q(s) \right. \\
&+ \left. \left( s^2+\omega^2 \right) \left( \alpha s + \beta \omega \right) \frac{\mathrm{d}q(s)}{\mathrm{d}s} \right\rbrace^{-1} .
\label{cont_trans_f3}
\end{aligned}
\end{equation}
In the limit $N \rightarrow +\infty $, the function $q(s)$ can be written as a hyperbolic sine, $\lim_{N \rightarrow + \infty} \frac{q(s)}{\omega^{2N}(N!)^2} =\frac{\omega}{\pi s}\sinh\left(\frac{\pi s}{\omega}\right)$, therefore the last expression reads (we divided the numerator and the denominator by the factor $\sinh\left(\pi s /\omega\right)$)
\begin{equation}
\begin{aligned}
C(s) =& -2\gamma \left( \alpha s^3 + \beta \omega s^2 \right) \\
&\times\left\lbrace \left( s^4 -\alpha s^3 +\omega (\omega-2\beta) s^2 + \alpha \omega^2 s \right) \right. \\
&+ \left( \alpha s^3+\beta \omega s^2+\alpha \omega^2 s+\omega^3 \beta \right)\\
&\times \left. \left[ \frac{\pi s}{\omega}\coth\left( \frac{\pi s}{\omega} \right)-1 \right]  \right\rbrace^{-1} .
\label{cont_trans_f4}
\end{aligned}
\end{equation}
Since the hyperbolic cotangent can be written as
\begin{equation}
\coth\left(\frac{\pi s}{\omega}\right)=\frac{1+\mathrm{e}^{-Ts}}{1-\mathrm{e}^{-Ts}},
\label{hcot}
\end{equation}
we end up with the following expression
\begin{equation}
\begin{aligned}
C(s) =& -2\gamma \left( \alpha s^3 + \beta \omega s^2 \right) \left[ 1-\mathrm{e}^{-Ts} \right] \\
&\times \left\lbrace s^4 \left(1+\frac{\alpha \pi}{\omega} \right) \left[ 1- \frac{\omega-\alpha\pi}{\omega+\alpha\pi} \mathrm{e}^{-Ts} \right] \right. \\
&+ s^3 (\beta\pi-2\alpha) \left[ 1- \frac{2\alpha+\beta\pi}{2\alpha-\beta\pi} \mathrm{e}^{-Ts} \right] \\
&+ s^2 \omega(\omega-3\beta+\alpha\pi)\left[ 1- \frac{\omega-3\beta-\alpha\pi}{\omega-3\beta+\alpha\pi} \mathrm{e}^{-Ts} \right] \\
&+ \left. s\beta\pi\omega^2 \left[ 1+\mathrm{e}^{-Ts} \right] -\beta\omega^3 \left[ 1-\mathrm{e}^{-Ts} \right] \right\rbrace^{-1} .
\label{cont_trans_f5}
\end{aligned}
\end{equation}
The inverse Laplace transformation of the last expression is rather tedious and technical work. Therefore, we moved it to the Appendix~\ref{app:inv_lap}. As a result, we obtain delay differential algebraic equations consisting of 4 first order differential equations
\begin{subequations}
\label{ndde_dqdt}
\begin{align}
\dot{q}_0(t) &= q_1(t)+f_0(t), \label{ndde_dqdt1} \\
\dot{q}_1(t) &= q_2(t)+f_1(t), \label{ndde_dqdt2} \\
\dot{q}_2(t) &= q_3(t)+f_2(t), \label{ndde_dqdt3} \\
\dot{q}_3(t) &= q_4(t)+f_3(t), \label{ndde_dqdt4}
\end{align}
\end{subequations}
5 algebraic equations with delayed terms
\begin{subequations}
\label{ndde_f}
\begin{align}
f_0(t) =& R_4 f_0(t-T) \nonumber \\
&+N_1 \left[ x(t-\tau)-x(t-\tau-T) \right], \label{ndde_f1} \\
f_1(t) =& R_4 f_1(t-T) \nonumber \\
&+N_2 \left[ x(t-\tau)-x(t-\tau-T) \right] \nonumber \\
&+M_3 \left[ f_0(t)-R_3 f_0(t-T) \right], \label{ndde_f2} \\
f_2(t) =& R_4 f_2(t-T) \nonumber \\
&+M_3 \left[ f_1(t)-R_3 f_1(t-T) \right] \nonumber \\
&+M_2 \left[ f_0(t)-R_2 f_0(t-T) \right], \label{ndde_f3} \\
f_3(t) =& R_4 f_3(t-T) \nonumber \\
&+M_3 \left[ f_2(t)-R_3 f_2(t-T) \right] \nonumber \\
&+M_2 \left[ f_1(t)-R_2 f_1(t-T) \right] \nonumber \\
&-M_1 \left[ f_0(t)+f_0(t-T) \right], \label{ndde_f4} \\
q_4(t) =& R_4 q_4(t-T) \nonumber \\
&+M_3\left[ q_3(t)-R_3 q_3(t-T) \right] \nonumber \\
&+ M_2 \left[ q_2(t)-R_2 q_2(t-T) \right] \nonumber \\
&- M_1 \left[ q_1(t)+ q_1(t-T) \right] \nonumber \\
&+ M_0 \left[ q_0(t)- q_0(t-T) \right], \label{ndde_f5}
\end{align}
\end{subequations}
and the set of parameters $R_2=\frac{\omega-3\beta-\alpha\pi}{\omega-3\beta+\alpha\pi}$, $R_3=\frac{2\alpha+\beta\pi}{2\alpha-\beta\pi}$, $R_4=\frac{\omega-\alpha\pi}{\omega+\alpha\pi}$, $N_1=\frac{2\alpha\omega}{\omega+\alpha\pi}$, $N_2=\frac{2\beta\omega^2}{\omega+\alpha\pi}$, $M_0=\frac{\beta\omega^4}{\omega+\alpha\pi}$, $M_1=\frac{\beta\pi\omega^3}{\omega+\alpha\pi}$, $M_2=\frac{\omega^2(3\beta-\alpha\pi-\omega)}{\omega+\alpha\pi}$ and $M_3=\frac{\omega(2\alpha-\beta\pi)}{\omega+\alpha\pi}$. The controller's output is $u(t)=-\gamma q_0(t)$. Since the controller as the input signal receives the delayed plant's output $x(t-\tau)$, the term $x(t-\tau-T)$ is the same delayed plant's output but additionally delayed by the period $T$. The controller~(\ref{ndde_dqdt}), (\ref{ndde_f}) produce absolutely the same output $u(t)$ as the controller~(\ref{all_harm}) for $N\rightarrow +\infty$ if both initial conditions are matched. Typical initial conditions for Eqs.~(\ref{all_harm}) is when all dynamical variables are set to zero: $a_j(0)=0$, $b_j(0)=0$. It corresponds to zero initial conditions for (\ref{ndde_dqdt}), (\ref{ndde_f}) when all dynamical variables $q_i$ and $f_i$ and their delay lines are set to zero. The important thing is that the delay line for $x(t-\tau)$ is also should be set to zero, meaning that the term $x(t-\tau-T)$ actually should be calculated as $x(t-\tau-T)\sigma(t-T)$ where $\sigma(t)$ stands for Heaviside step function.

While it might look like the controller~(\ref{ndde_dqdt}), (\ref{ndde_f}) is superior to the truncated version~(\ref{all_harm}) since it deals with all harmonics, that is not exactly true. The involvement of all harmonics is an advantage and a disadvantage at the same time. In a typical experiment, the signal $x(t-\tau)$ is measured not continuously but at discrete time moments with a fixed time step. It means that Eqs.~(\ref{ndde_dqdt}), (\ref{ndde_f}) also should be integrated using a fixed time step integration scheme. Since the controller~(\ref{ndde_dqdt}), (\ref{ndde_f}) effectively contains all harmonics, the fixed time step integration scheme applied to~(\ref{ndde_dqdt}), (\ref{ndde_f}) should deal with extremely high harmonic numbers. But it means that we have a situation when the period of the harmonic oscillator (with extremely high harmonic number) is much lower than the integration step. Therefore, it leads to an inaccuracy of the integration scheme. However, such inaccuracy is not visible at a short time interval, and only at a long time interval does the error accumulate and give unstable dynamics of the controller. In contrast, the truncated controller~(\ref{all_harm}) does not have such an issue.

\begin{figure}[h!]
\centering\includegraphics[width=0.95\columnwidth]{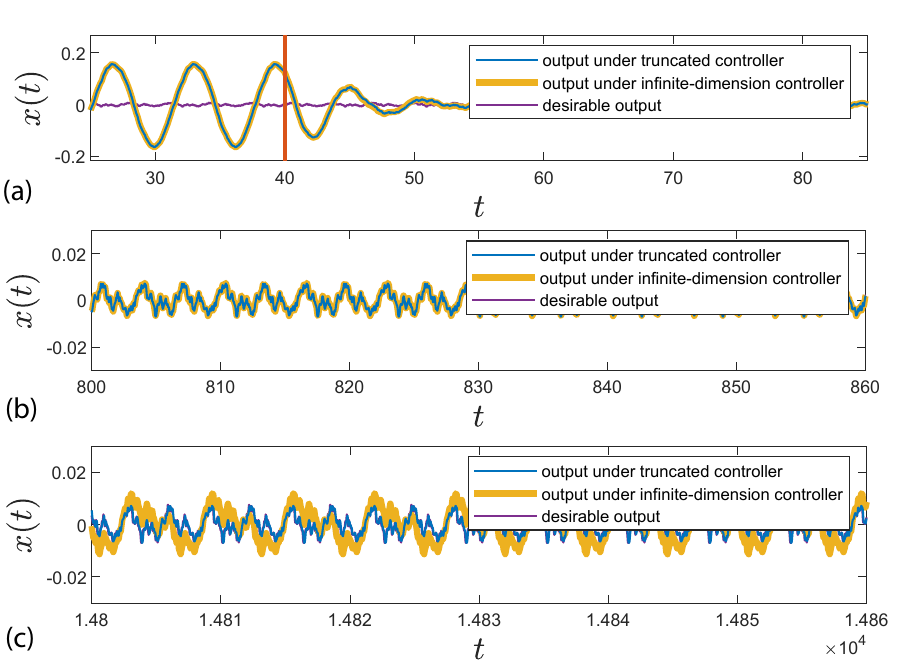}
\caption{\label{fig7} The comparison of both controllers: the controller~(\ref{ndde_dqdt}), (\ref{ndde_f}) and truncated controller~(\ref{all_harm}). Similar to Figure~\ref{fig4}, the blue line represents the truncated controller, and the purple line is the external force without the first harmonic (desirable output). Yet, here we have a thick yellow line representing the controller~(\ref{ndde_dqdt}), (\ref{ndde_f}). The parameters are the same as in Figure~\ref{fig4} except now the external force has 80 harmonics, and the truncated controller has only 5 harmonics. One can see that in (a) and (b), the blue line coincides with the thick yellow line, but after a long time interval in (c), the yellow line alternates from the desirable output.}
\end{figure}

To compare the truncated controller with the controller described by the delay differential algebraic equations we performed following computation: the plant equation~(\ref{main}) is integrated using high-precision adaptive time step method $\mathtt{ode45}$ (standard MatLab integrator) while the controller Eqs.~(\ref{all_harm}) or Eqs.~(\ref{ndde_dqdt}), (\ref{ndde_f}) is integrated using the fixed time step Adams-Bashforth 3rd order method. In Figure~\ref{fig7}, we depicted the successful work of both controllers on different time scales. Figure~\ref{fig7}(a) shows transient dynamics when controllers are turned on. Both dynamics almost coincide. Figure~\ref{fig7}(b) shows the dynamics after some time of working controllers. Again, both dynamics almost coincide. In contrast, Figure~\ref{fig7}(c) depicts the dynamics after a long time of working controllers. Here, one can see that accumulated error gives increasing amplitude for the controller~(\ref{ndde_dqdt}), (\ref{ndde_f}) while the truncated controller remains stable.

\section{\label{sec:conc}Conclusions}

We study the situation in which the plant is disturbed by a periodic signal of a known frequency. Additionally, there is a delay line between the plant and controller. The delay time is assumed to be unknown. We developed a control algorithm to eliminate the first harmonic in the output of the plant without disturbing all the other harmonics in the periodic signal. The algorithm is based on a modified version of the system of harmonic oscillators~\cite{Olyaei2015,AzimiOlyaei2018} used to stabilize an unstable periodic orbit with an unknown profile. The demand for such an algorithm naturally appears in the experiments of rotational scanning AFM~\cite{Ulcinas2017}, and it has been successfully realized in a recent experiment~\cite{VAITEKONIS2025117552}.

Our controller is a linear time-invariant system described by the transfer function~(\ref{cont_trans_f2}). This transfer function has three undefined constants: the coupling constants $\alpha$ and $\beta$, and the gain factor $\gamma$. We calculated the controller's stability (see Figure~\ref{fig3}) for different time delays and observed that a good choice of the coupling constants is $\alpha=0.1$, $\beta=-0.1$. The factor $\gamma$ is not precisely the controller's gain $K$ which appears in the controller's equations~(\ref{all_harm}) and (\ref{out}). Because we used a particular form of the coupling constants~(\ref{cons_form}), an interchange of $K$ to $\gamma$ occurred. This means that in practical realization, we should guess the value $\gamma$ and set $K=\gamma$. The parameter $\gamma$ has a well-defined physical interpretation: the relaxation time of the plant variable or the stiffness of the cantilever in the case of AFM. Therefore, some information on $\gamma$ can be obtained a priori. However, an inaccurate estimate of $\gamma$, in the worst-case scenario can only lead to instability. Meaning that it does not change the purpose of the controller, that is, to eliminate the first harmonic.

The controller can be extended in various directions. For example, one can ask whether it is possible to eliminate not only the first harmonic but also, similar to~\cite{Mattavelli2004}, some prescribed harmonics. Intuitively, due to the time delay, the more harmonics we want to eliminate, the more unstable is the controller. Yet, this is an unexplored possibility that require further investigation. Another direction could be the extension of the controller to operate with unknown frequencies. Developments in this direction~\cite{Bodson1997,Bodson2016} can help solve this problem. Another potentially interesting question is how to deal with longer delay times. When we know that the delay is not smaller than some particular value, a classical Smith predictor~\cite{Smith57} together with our proposed controller can manage such a task in the following way: the predictor compensates for the large known part of the dead time, while our controller deals with the remaining uncertain delay. Since the plant transfer function~(\ref{plant_trf}) is extremely simple, the classical Smith predictor or its modification~\cite{Zhou2007,P.PrincesSindhuja2023} should fit well for this task. However, additional research is required, which can be done in future work.

\section*{Data availability}

Data sharing is not applicable to this article as no new data were created or analyzed in this study.

\section*{Funding}

No funding received.

\section*{Acknowledgements}

We thank E.~Anisimovas for advice on the transfer function stability criteria, and A.~Kononovi\v{c}ius for comments on numerical simulation of stochastic differential equations.

\section*{Author contributions statement}

V.N. worked on the idea, wrote the article, conducted the analysis and simulations, Š.V. performed visualization and analyzed results. All authors reviewed the manuscript.

\section*{Pre-print of the manuscript}

The pre-print version of this manuscript is available at~\cite{arxiv2403.11731,authorea2024}.

\section*{Competing interests}

Viktor Novičenko has patent pending to Center for Physical Sciences and Technology, and Vilnius University. Šarūnas Vaitekonis has patent pending to Center for Physical Sciences and Technology, and Vilnius University.

\bibliography{references}

\appendix

\section{\label{app:trans}Derivation of the controller transfer function}

Here we will derive the transfer function for the controller in the main text defined as a ratio of the form $C(s) = \allowbreak U(s)/[X(s)\mathrm{e}^{-\tau s}] \allowbreak =  K A_1(s)/[X(s)\mathrm{e}^{-\tau s}]$.

First, let us introduce complex-valued dynamical variables $\tilde{a}_j(t)$ defined as $\tilde{a}_j(t)=[a_j(t)+\mathrm{i}b_j(t)]/2$ for $j=1,2,\ldots,N$, $\tilde{a}_j(t)=[a_{-j}(t)-\mathrm{i}b_{-j}(t)]/2$ for $j=-1,-2,\ldots,-N$ and $\tilde{a}_0(t)=a_0(t)$. The differential equations for them have more symmetry. Indeed, the original paper on the MHO~\cite{Olyaei2015} operates in this setup. Hence, the main text Eqs.~\eqref{all_harm} read
\begin{equation}
\dot{\tilde{a}}_j(t) = \mathrm{i} j \omega \tilde{a}_j(t)+\kappa_j \left[ x(t-\tau) - \sum_{i=-N \atop |i| \neq 1}^{N} \tilde{a}_i(t) \right],
\label{all_harm_compl}
\end{equation}
where $j=-N,-N+1,\ldots,N$. Here we introduced new coupling constants $\kappa_j=[\alpha_j+\mathrm{i}\beta_j]/2$ for positive $j$, $\kappa_j=[\alpha_{-j}-\mathrm{i}\beta_{-j}]/2$ for negative $j$, and $\kappa_0=\alpha_0$. Therefore, we have $\tilde{a}_j^{*}(t)=\tilde{a}_{-j}(t)$ and $\kappa_j^{*}=\kappa_{-j}$ meaning that the complex conjugation is equivalent to the sign flipping of the index.

The next step is to apply the Laplace transformation for the system of differential equations~(\ref{all_harm_compl}). By introducing the column vector $\tilde{\mathbf{A}}(s)=(\tilde{A}_{-N}(s),\tilde{A}_{-N+1}(s),\ldots,\tilde{A}_{N}(s))^T$ (here $^T$ denotes transposition) containing the Laplace transformations for all dynamical variables $\tilde{a}_j(t)$, one can write~(\ref{all_harm_compl}) as
\begin{equation}
\mathbf{M} \tilde{\mathbf{A}}(s) = \mathrm{e}^{-\tau s} X(s) \left(
\begin{array}{c}
\kappa_{-N} \\
\kappa_{-N+1} \\
\vdots \\
\kappa_{N}
\end{array}
\right).
\label{lapl_all}
\end{equation}
The matrix $\mathbf{M}$ has following form
\begin{equation}
\begin{aligned}
\mathbf{M} =& \mathrm{diag}\left[ s-\mathrm{i}(-N)\omega, s-\mathrm{i}(-N+1)\omega,\ldots,s-\mathrm{i}N\omega \right] \\
&+ \left(
\begin{array}{c}
\kappa_{-N} \\
\kappa_{-N+1} \\
\vdots \\
\kappa_{N}
\end{array}
\right) \left(
\begin{array}{ccccccccc}
1 & \ldots & 1 & 0 & 1 & 0 & 1 & \ldots & 1
\end{array}
\right),
\label{mat_m}
\end{aligned}
\end{equation}
where $\mathrm{diag}[c_1,c_2,\ldots]$ denotes a diagonal matrix with the elements $c_1$, $c_2$, $\ldots$ on the diagonal. The matrix $\mathbf{M}$ has a special structure: it is written as a sum of the diagonal matrix and the outer product matrix. Moreover, one of the vectors in the construction of the outer product matrix has only ones and zeros. Because of such a special form, one can easily find an inverse matrix. Using notation $p_j(s)=1/(s-\mathrm{i}j\omega)$ one can read
\begin{equation}
\begin{aligned}
\mathbf{M}^{-1} = &\mathrm{diag}\left[ p_{-N} , p_{-N+1} ,\ldots, p_{N} \right]  \\
&-\left[ 1+ \sum_{j=-N \atop |j| \neq 1}^{N} \kappa_j p_j \right]^{-1} \left(
\begin{array}{c}
\kappa_{-N}p_{-N} \\
\kappa_{-N+1} p_{-N+1} \\
\vdots \\
\kappa_{N} p_N
\end{array}
\right) \\
& \times \left(
\begin{array}{ccccccccc}
p_{-N} & \ldots & p_{-2} & 0 & p_0 & 0 & p_2 & \ldots & p_N
\end{array}
\right).
\label{mat_invm}
\end{aligned}
\end{equation}
By combining Eqs.~(\ref{mat_invm}) and (\ref{lapl_all}), we finally obtain the Laplace transform of the output signal
\begin{equation}
\begin{aligned}
U(s) =& K \left( \tilde{A}_{-1}(s)+\tilde{A}_{1}(s) \right) = K \mathrm{e}^{-\tau s} X(s)\\
 \times&  \left[ 1+ \sum_{j=-N \atop |j| \neq 1}^{N} \frac{\kappa_j}{s-\mathrm{i}j\omega}  \right]^{-1} \left( \frac{\kappa_{-1}}{s+\mathrm{i}\omega}+\frac{\kappa_{1}}{s-\mathrm{i}\omega} \right).
\label{out_lapl}
\end{aligned}
\end{equation}
Subsequently, the transfer function
\begin{equation}
C(s) = K \left[ 1+ \frac{\alpha_0}{s} + \sum_{j=2}^{N} \frac{\alpha_j s- j \beta_j \omega}{s^2+ j^2\omega^2}  \right]^{-1} \left( \frac{\alpha_1 s-\beta_1 \omega}{s^2+\omega^2} \right).
\label{cont_trans_f1}
\end{equation}

\section{\label{app:stab}Stability condition for the closed-loop system}

The polynomial~\eqref{poles3} in the main text is the $(2N+1)$-order polynomial with real coefficients. The necessary condition for the stability is that all coefficients should be positive. The coefficient next to the term $s^{2N}$ is $\alpha(2N+1)$. Thus, the necessary stability condition is
\begin{equation}
\alpha>0.
\label{nec_cond1}
\end{equation}
When $\alpha$ is positive, the only way to lose (or gain) stability is the case when the pair (or several pairs) of complex conjugate roots cross the imaginary line $\Re(s)=0$, because $s=0$ can not be a solution of the main text Eq.~\eqref{poles3} for $\alpha \neq 0$.

Next, let us look for the coefficient next to the term $s^1$:
\begin{equation}
\omega^{2N}\left( N! \right)^2 \left[ 1+2\frac{\beta}{\omega} \sum_{k=1}^N \frac{1}{k^2} \right].
\label{coef_s}
\end{equation}
From last, we obtain the second necessary stability condition (we take the limit $N\rightarrow + \infty$)
\begin{equation}
\frac{\beta}{\omega}>-\frac{3}{\pi^2}.
\label{nec_cond}
\end{equation}
We emphasize that such a condition is only necessary but not sufficient; thus, it does not give the border of the stability region.

Now, we can analyze how the roots move when we change the parameters $(\alpha,\beta)$ from the point $(\alpha,\beta)=(0,0)$ to any other relevant point. Note that the point is relevant if it is in the region $\alpha>0$ and $\beta/\omega>-3/\pi^2$ since all other points are proved to be unstable; therefore, the important consequence is that any relevant point can be achieved without crossing the line $\alpha=0$. At the point $(0,0)$ all roots are on the imaginary axis, $s_k=\mathrm{i}k\omega$ with $k=-N, -N+1,\ldots, N-1,N$. The roots $s_k(\alpha,\beta)$ are continuous functions on the parameters $(\alpha,\beta)$, and to find how the roots move by slightly changing $(\alpha,\beta)$ one should find an exact differential $\mathrm{d}s_k(\alpha,\beta)$ at the point $(0,0)$. In the limit $N \rightarrow +\infty$ the main text equation~\eqref{poles3} simplifies to
\begin{equation}
\begin{aligned}
&p(s,\alpha,\beta) = (s+\alpha) \frac{\omega}{\pi s} \sinh\left( \frac{\pi s}{\omega} \right) \\
&+ \left(\alpha s + \beta \omega \right) \left[ \frac{1}{s} \cosh\left( \frac{\pi s}{\omega}\right)-\frac{\omega}{\pi s^2} \sinh\left( \frac{\pi s}{\omega}\right) \right] =0,
\label{poles4}
\end{aligned}
\end{equation}
where we have used the identity
\begin{equation}
\lim_{N\rightarrow +\infty} \frac{q(s)}{\omega^{2N}\left( N! \right)^2}=\frac{\omega}{\pi s}\sinh\left( \frac{\pi s}{\omega} \right).
\label{sinh_lim}
\end{equation}
Thus, the exact differential become
\begin{equation}
\begin{aligned}
\mathrm{d}s_k(0,0) =& - \left. \frac{\frac{\partial p(s,\alpha,\beta)}{ \partial \alpha }}{ \frac{\partial p(s,\alpha,\beta)}{\partial s}} \right|_{(s_k,0,0)} \mathrm{d}\alpha - \left. \frac{\frac{\partial p(s,\alpha,\beta)}{ \partial \beta }}{ \frac{\partial p(s,\alpha,\beta)}{\partial s}} \right|_{(s_k,0,0)} \mathrm{d}\beta \\
=& (-1) \cdot \mathrm{d}\alpha + \mathrm{d}\beta \cdot
\begin{cases}
 \frac{\mathrm{i}}{k}  & \mathrm{for} \; k\neq 0 \\
0 & \mathrm{for} \; k=0 
\end{cases}
.
\label{exactd}
\end{aligned}
\end{equation}
Since the factor next to $\mathrm{d}\alpha$ is purely real (negative) and the factor next to $\mathrm{d}\beta$ is purely imaginary, the small step from $(\alpha,\beta)=(0,0)$ to $(\alpha,\beta)=(\mathrm{d}\alpha,\mathrm{d}\beta)$ with positive $\mathrm{d}\alpha$ and any $\mathrm{d}\beta$ would move all roots to the left half-plane (stable region). If the system loses stability and we do not need to cross the line $\alpha=0$, one should observe when the pair (or several pairs) of complex conjugate roots cross the imaginary axis. Such event can happen only when $p(\mathrm{i}\omega\eta,\alpha,\beta)=0$ for real $\eta\neq0$ giving
\begin{equation}
\begin{aligned}
\alpha \cos(\pi\eta) &=0, \\
\frac{\omega}{\pi} \left[ 1+\frac{\beta}{\omega \eta^2} \right]\sin(\pi \eta) -\frac{\beta}{\eta}\cos(\pi\eta)&=0.
\label{eta_cond}
\end{aligned}
\end{equation}
From the first equation we obtain that $\eta \in \big\{ \pm 1/2, \pm 3/2, \allowbreak \pm 5/2, \ldots \big\}$. According to that, the second equation gives $-\beta/\omega \allowbreak \in  \big\{ 1/4, \allowbreak 9/4, 25/4, \ldots \big\}$. Thus, up to $\beta/\omega>-1/4$, the system remains stable and loses its stability at $\beta/\omega=-1/4$. It potentially can gain stability at $\beta/\omega=-9/4$, but since Eq.~(\ref{nec_cond}) guarantee the instability for $\beta/\omega<-3/\pi^2$ we end up with the necessary and sufficient condition for the stability of the closed-loop system: $\alpha>0$ and $\beta/\omega>-1/4$.

\section{\label{app:inv_lap}Inverse Laplace transform of the transfer function~\eqref{cont_trans_f5} of the main text}

Here, we will derive controller equations in the time domain when the transfer function is defined by the main text Eq.~\eqref{cont_trans_f5}. To shorten notations, we will assume that the controller's input signal is $x(t)$ instead of $x(t-\tau)$. Generalization to the case $\tau \neq 0$ is straightforward.

According to definition $C(s)=U(s)/X(s)$ and using notation $U(s)=-\gamma Q_0(s)$ we obtain
\begin{equation}
\begin{aligned}
& 2 \alpha s^3 X(s)\left[ 1-\mathrm{e}^{-Ts} \right] + 2\beta \omega s^2 X(s) \left[ 1-\mathrm{e}^{-Ts} \right] = \\
&  \frac{\omega+\alpha \pi}{\omega} s^4 Q_0(s) \left[ 1- R_4 \mathrm{e}^{-Ts} \right]   \\
&+ (\beta\pi-2\alpha) s^3 Q_0(s) \left[ 1- R_3 \mathrm{e}^{-Ts} \right] \\
&+ \omega(\omega-3\beta+\alpha\pi) s^2 Q_0(s) \left[ 1- R_2 \mathrm{e}^{-Ts} \right] \\
&+  \beta\pi\omega^2 sQ_0(s) \left[ 1+\mathrm{e}^{-Ts} \right] - \beta\omega^3 Q_0(s) \left[ 1-\mathrm{e}^{-Ts} \right],
\label{inv_lap}
\end{aligned}
\end{equation}
where we used notation $R_2=\frac{\omega-3\beta-\alpha\pi}{\omega-3\beta+\alpha\pi}$, $R_3=\frac{2\alpha+\beta\pi}{2\alpha-\beta\pi}$ and $R_4=\frac{\omega-\alpha\pi}{\omega+\alpha\pi}$. Let us introduce new variables $Q_{1,2,3,4}(s)$ and $P_{0,1,2,3}(s)$ and relations for them
\begin{subequations}
\label{qp}
\begin{align}
s Q_0(s) &= Q_1(s)+P_0(s)X(s), \label{qp1} \\
s Q_1(s) &= Q_2(s)+P_1(s)X(s), \label{qp2} \\
s Q_2(s) &= Q_3(s)+P_2(s)X(s), \label{qp3} \\
s Q_3(s) &= Q_4(s)+P_3(s)X(s). \label{qp4}
\end{align}
\end{subequations}
From last we obtain expressions for $s^2Q_0(s)$, $s^3Q_0(s)$ and $s^4Q_0(s)$:
\begin{subequations}
\label{qs}
\begin{align}
s^2 Q_0(s) &= Q_2(s)+P_1(s)X(s)+P_0(s)sX(s), \label{qs1} \\
s^3 Q_0(s) &= Q_3(s)+P_2(s)X(s)+P_1(s)sX(s) \nonumber \\
&+P_0(s)s^2 X(s), \label{qs2} \\
s^4 Q_0(s) &= Q_4(s)+P_3(s)X(s)+P_2(s)sX(s) \nonumber \\
&+P_1(s)s^2 X(s) + P_0(s)s^3 X(s). \label{qs3}
\end{align}
\end{subequations}
Substitution of the last expressions into Eq.~(\ref{inv_lap}) and collecting the terms next to $s^3 X(s)$ gives
\begin{equation}
P_0(s)= \frac{2\alpha \omega}{\omega+\alpha\pi} \frac{1-\mathrm{e}^{-Ts}}{1-R_4\mathrm{e}^{-Ts}},
\label{P0}
\end{equation}
next to $s^2 X(s)$ gives
\begin{equation}
\begin{aligned}
P_1(s) &= \frac{2\beta \omega^2}{\omega+\alpha\pi} \frac{1-\mathrm{e}^{-Ts}}{1-R_4\mathrm{e}^{-Ts}} \\
&+\frac{\omega(2\alpha-\beta\pi)}{\omega+\alpha\pi}\frac{1-R_3\mathrm{e}^{-Ts}}{1-R_4\mathrm{e}^{-Ts}} P_0(s),
\label{P1}
\end{aligned}
\end{equation}
next to $s X(s)$ gives
\begin{equation}
\begin{aligned}
P_2(s) &= \frac{\omega(2\alpha-\beta\pi)}{\omega+\alpha\pi} \frac{1-R_3\mathrm{e}^{-Ts}}{1-R_4\mathrm{e}^{-Ts}} P_1(s) \\
&+\frac{\omega^2 (3\beta-\alpha\pi-\omega)}{\omega+\alpha\pi}\frac{1-R_2\mathrm{e}^{-Ts}}{1-R_4\mathrm{e}^{-Ts}} P_0(s),
\label{P2}
\end{aligned}
\end{equation}
and finally next to $X(s)$ gives
\begin{equation}
\begin{aligned}
P_3(s) &= \frac{\omega(2\alpha-\beta\pi)}{\omega+\alpha\pi} \frac{1-R_3\mathrm{e}^{-Ts}}{1-R_4\mathrm{e}^{-Ts}} P_2(s) \\
&+\frac{\omega^2 (3\beta-\alpha\pi-\omega)}{\omega+\alpha\pi}\frac{1-R_2\mathrm{e}^{-Ts}}{1-R_4\mathrm{e}^{-Ts}} P_1(s) \\
&-\frac{\omega^3 \beta\pi}{\omega+\alpha\pi}\frac{1+\mathrm{e}^{-Ts}}{1-R_4\mathrm{e}^{-Ts}} P_0(s).
\label{P3}
\end{aligned}
\end{equation}
Using the notification $P_i(s)X(s)=F_i(s)$ and multiplying both sides of (\ref{P0}), (\ref{P1}), (\ref{P2}), (\ref{P3}) by $X(s)$ we obtain 4 algebraic equations for the variables $f_i(t)=\mathcal{L}^{-1}\left(F_i(s)\right)$:
\begin{subequations}
\label{f}
\begin{align}
f_0(t) =& R_4 f_0(t-T) \nonumber \\
&+N_1 \left[ x(t)-x(t-T) \right], \label{f1} \\
f_1(t) =& R_4 f_1(t-T) \nonumber \\
&+N_2 \left[ x(t)-x(t-T) \right] \nonumber \\
&+M_3 \left[ f_0(t)-R_3 f_0(t-T) \right], \label{f2} \\
f_2(t) =& R_4 f_2(t-T) \nonumber \\
&+M_3 \left[ f_1(t)-R_3 f_1(t-T) \right] \nonumber \\
&+M_2 \left[ f_0(t)-R_2 f_0(t-T) \right], \label{f3} \\
f_3(t) =& R_4 f_3(t-T) \nonumber \\
&+M_3 \left[ f_2(t)-R_3 f_2(t-T) \right] \nonumber \\
&+M_2 \left[ f_1(t)-R_2 f_1(t-T) \right] \nonumber \\
&-M_1 \left[ f_0(t)+f_0(t-T) \right], \label{f4}
\end{align}
\end{subequations}
where a set of new constants is introduced $N_1=\frac{2\alpha\omega}{\omega+\alpha\pi}$, $N_2=\frac{2\beta\omega^2}{\omega+\alpha\pi}$, $M_0=\frac{\beta\omega^4}{\omega+\alpha\pi}$, $M_1=\frac{\beta\pi\omega^3}{\omega+\alpha\pi}$, $M_2=\frac{\omega^2(3\beta-\alpha\pi-\omega)}{\omega+\alpha\pi}$ and $M_3=\frac{\omega(2\alpha-\beta\pi)}{\omega+\alpha\pi}$. The inverse Laplace transform of Eqs.~(\ref{qp}) gives 4 first order differential equations
\begin{subequations}
\label{app_dqdt}
\begin{align}
\dot{q}_0(t) &= q_1(t)+f_0(t), \label{app_dqdt1} \\
\dot{q}_1(t) &= q_2(t)+f_1(t), \label{app_dqdt2} \\
\dot{q}_2(t) &= q_3(t)+f_2(t), \label{app_dqdt3} \\
\dot{q}_3(t) &= q_4(t)+f_3(t). \label{app_dqdt4}
\end{align}
\end{subequations}
Finally the algebraic equation for $q_4(t)$ is obtained as a reminder of the substitution of (\ref{qs}) to (\ref{inv_lap}):
\begin{equation}
\begin{aligned}
q_4(t) &= R_4q_4(t-T)+M_3 \left[ q_3(t)-R_3q_3(t-T) \right] \\
&+M_2 \left[ q_2(t)-R_2q_2(t-T) \right]-M_1 \left[ q_1(t)+q_1(t-T) \right] \\
&+M_0\left[ q_0(t)-q_0(t-T) \right].
\label{qt4}
\end{aligned}
\end{equation}

\end{document}